\documentclass{article}
\usepackage{amssymb}
\usepackage{amstext}
\newcommand*\de{\mathrm{d}}
\newcommand*\De{\mathrm{D}}
\renewcommand*\epsilon{\varepsilon}
\renewcommand*\phi{\varphi}
\renewcommand*\theta{\vartheta}
\begin{document}
\title{\bf Second order formalism in Poincar\'e gauge theory} 
\author{ M. Leclerc \\ 
 \small Section of Astrophysics and Astronomy, Department of Physics, \\
 \small  University of Athens, Greece} 
\maketitle 
\begin{abstract}
Changing the set of independent variables of Poincar\'e gauge theory and 
considering, 
in a manner similar to the second order formalism of general relativity, 
the Riemannian part of the Lorentz 
 connection as function of the tetrad field, 
we   construct theories that do not contain second or higher order 
derivatives in the field variables, possess a full general relativity 
limit in the absence of spinning matter fields, and allow for 
propagating torsion fields in the general case, the spin density 
playing the role of the source current in a Yang-Mills type equation 
for the torsion. The equivalence of the second order and 
conventional first order formalism is  established and  
the corresponding Noether identities are discussed. Finally, a concrete 
Lagrangian is constructed and by means of a Yasskin type ansatz, the 
field equations are reduced to a conventional Einstein-Proca system. 
Neglecting higher order terms in the spin tensor, approximate solutions 
describing the exterior of a spin polarized neutron star are presented     
and the possibility of the experimental detection  of the torsion fields 
is briefly discussed.  

\end{abstract}

\section{Introduction}
In general relativity, the procedure to consider, during the variation 
of the Lagrangian, the connection 
and the metric as independent of each other 
is known as Palatini, or \textit{first order}, formalism. Indeed 
 the classical Hilbert-Einstein Lagrangian contains no second derivatives 
of those fields. On the other hand, in the conventional metric approach 
of general relativity, where the only independent field is the metric, 
the Lagrangian 
 contains second derivatives of the metric, although incidentally, the 
terms containing them can be eliminated by subtracting a boundary term. 
This is known as \textit{second order} formalism. 

In Poincar\'e gauge theory (PGT), the usual procedure is to consider 
the connection $\Gamma^{ab}_{\ \ m}$ and the tetrad $e^a_m$ as 
independent fields and consequently, the field equations are 
derived through variation with respect to those fields. For a detailed 
description of PGT, we refer to Refs. \cite{1} and \cite{2}.
Usually, 
the Lagrangians 
are constructed from terms at most second order in curvature and 
torsion, and thus contain no second  derivatives of the 
independent fields. We will therefore refer to this as the first 
order formalism.  

However, in view of the 
relation $\Gamma^{ab}_{\ \ m} = \hat\Gamma^{ab}_{\ \ m} 
+ K^{ab}_{\ \ m}$, where $\hat \Gamma^{ab}_{\ \ m}$ is the Christoffel 
connection, which  can be expressed as 
function of the tetrad (and its derivatives) only, and $K^{ab}_{\ \ m}$ 
is the contortion tensor, one might  
equally well consider $K^{ab}_{\ \ m}$ and $e^a_m$ as independent fields 
and carry out the variation with respect to those fields. Then, similar  
to the classical approach in  general relativity (GR), 
the Christoffel part of the connection is considered 
as a function of the tetrad. 
We will  refer to this as the second order formalism.

What could be the use of this formalism? Well, in conventional PGT, the 
most general Lagrangian is constructed under the requirement that 
it should not contain second and higher order derivatives of 
the independent fields 
$e^a_m$ and $\Gamma^{ab}_{\ \ m}$ for the usual reasons (Cauchy problem). 
It is easy to see that, if we require instead the Lagrangian not to  
 contain second and higher order derivatives of $e^a_m$ and 
$K^{ab}_{\ \  m}$, now seen as independent fields, then this will 
allow for other terms in the Lagrangian. 
Also, some terms previously allowed are now forbidden, as is the case, e.g., 
 for 
terms of second order in the curvature, which will contain second derivatives 
of the tetrad field in this formalism. 
The Cauchy problem can now be 
stated with respect to $e^a_m$ and $K^{ab}_{\ \ m}$. Since 
the determination of the latter fields allows for the determination of 
$\Gamma^{ab}_{\ \ m}$ too,  no initial value problem will arise this way. 

On the other hand, one could argue that, in view of the underlying gauge 
structure of the theory, it appears quite unnatural to consider 
as fundamental variables a set different to the gauge fields of the 
Poincar\'e group. It turns out, however, that both formalisms are 
completely equivalent. In other words, it is simply a matter of choice
 and convenience which formalism we  use. 
This equivalence ultimately means that
 theories that contain higher derivatives in one formalism may 
be reformulated in the other formalism such that no higher derivatives 
occur. The change of the fundamental variables thus  does not change 
anything concerning  
the viability or consistency of the theory, but it shows that 
certain theories that are apparently in conflict with the Cauchy initial value 
problem need not be so in reality.

Summarizing, the second order formalism allows us to investigate  different 
Lagrangians than those usually considered in PGT. This, in turn, will 
allow us to construct a theory with a complete classical general 
relativity limit, and thus a complete agreement with the experimental 
situation, without any constraints on eventual parameters of the theory, 
but with dynamical torsion fields arising in the presence of spinning 
matter. As we have pointed out in Ref. \cite{3}, this is not possible 
in conventional PGT for the following reasons. Since in the usual 
first order approach, the independent fields are the tetrad $e^a_m$ and 
the Lorentz connection $\Gamma^{ab}_{\ \ m}$, in order to get  
propagating modes for both fields, their first derivatives should appear 
quadratically in the Lagrangian. Therefore, apart from a term suitable
 of producing the general relativity limit, there has to be at least 
one term quadratic in the curvature tensor. In the presence of spinless  
matter, we wish the field equations to reduce to the Einstein equations 
of general relativity. If the Cartan equation (i.e., the connection equation)
 in this case leads to 
a vanishing torsion (Riemannian geometry), then the term quadratic in the 
curvature  
 will necessarily contribute to the Einstein equation, which will 
thus not be of the GR form. (And therefore, in order to be in 
agreement with experiments, constraints will have to be imposed on 
the coupling constants of such theories.) 
The only exceptions to this are very 
artificial Lagrangians, with terms of the form $R_{[ik]}R^{[ik]}$ and 
$R^l_{\ [ikm]}R_l^{\ [ikm]}$,  constructed from the antisymmetric   
 part of the Ricci tensor and the completely antisymmetric (in the last three 
indices) part of the curvature tensor. Those terms will lead to contributions 
in the Einstein equation that vanish together with the torsion in 
the spinless case (in view  of the Bianchi identities in the 
Riemannian limit). In the presence of spinning matter fields, those terms 
will eventually lead to propagating torsion, but in general, the field 
equations have very few and quite strange solutions, certainly not of 
the form expected for a Yang-Mills like theory. On the other hand, 
theories whose 
 Cartan equation in the vanishing spin limit leads to a vanishing 
curvature (teleparallel geometry) and presenting a 
 Yang-Mills like behavior for the Lorentz connection can easily be 
constructed using curvature 
squared terms, without problems with the general relativity limit. 
Those theories have been analyzed in detail in Ref. \cite{3}, where we have 
shown that they all present problems with an additional symmetry arising 
in the classical limit. Schematically, we can write the linear 
approximation of such theories in the following form of wave equations   
\begin{equation}
\Box \Gamma^{ab}_{\ \ m} = \sigma^{ab}_{\ \ m}\ \  \text{and}\ \ 
 \Box g_{ik} = T_{ik} + \mathcal{O}(R), 
\end{equation}
which leads, in the spinless case, to $\Gamma^{ab}_{\ \ m} = 0$, and 
therefore the curvature corrections $\mathcal{O}(R)$ 
will vanish and we are left 
with the GR equation $\Box g_{ik} = T_{ik}$. The problem is that, even 
for $\Gamma^{ab}_{\ \ m}$ fixed to zero, we  can Lorentz rotate the 
tetrad field $e^a_m \rightarrow \Lambda^a_{\ b} e^b_m$, without 
changing the metric $g_{ik}$. Thus, the tetrad field is not uniquely 
fixed by the field equations, and therefore, the behavior of 
spinning test matter, 
e.g., the spin precession and the trajectory of a Dirac particle, cannot 
be uniquely predicted \cite{3}. It is clear that this 
argumentation is valid independently of the formalism (first or second) 
one uses for the derivation of the field equations. It is a general 
problem of the set of equations 
$\Gamma^{ab}_{\ \ m}=0$ and $G_{ik} = T_{ik}$, i.e., of the teleparallel 
form of GR. 

On the other hand, a theory with an GR limit in a Riemannian geometry 
(vanishing torsion), should behave as 
\begin{equation}
\Box K^{ab}_{\ \ m} = \sigma^{ab}_{\ \ m}\ \  \text{and}\ \ 
 \Box g_{ik} = T_{ik} + \mathcal{O}(K), 
\end{equation}
 where now the spinless case leads to a vanishing contortion $K^{ab}_{\ \
 m}=0$, and thus to a vanishing torsion, and therefore, as before, 
 the second equation reduces 
to the field equations of GR. In contrast to (1), the set of equations 
(2) fixes the geometry completely. Indeed, if we know the contortion and 
the metric, than the independent field tensors, torsion and curvature, 
are fixed, 
up to a Poincar\'e transformation of course, which is the underlying 
gauge symmetry of the theory \cite{3}. 

However, as we have argued above, no Lagrangian of conventional PGT 
leads to equations of the form (2). In order to get such field equations 
using the conventional first order approach, one would have to 
use higher order derivatives of the fields $(\Gamma^{ab}_{\ \ m}, e^a_m )$
in the Lagrangian, which is 
(apparently) forbidden in view of the initial value problem.  

We will solve this problem using the second order approach and construct 
an explicit example of a theory with equations of type (2), containing 
only first order derivatives of the independent fields $(K^{ab}_{\ \ m},
e^a_m)$.

The article is structured as follows. 
In the next section, we will establish the equivalence of the first and 
second order formalism as far as the field equations are concerned. 
In section 3, we will compare the Noether currents and their  
conservation equations in the two approaches. 
Section 4 is devoted to the construction of an exemplifying  theory that, 
using 
the second order formalism, allows for a full general relativity limit 
in the absence of spinning matter. Dynamical torsion fields will arise 
if the matter possesses a non-vanishing   spin density.   
In section 5, extending  slightly the Yasskin ansatz known 
from Einstein-Yang-Mills theory with internal symmetry group, 
we simplify the system of field equations for a special 
class of solutions, suitable for the description of spin polarized 
neutron stars. Finally, in section 6, explicit solutions are discussed 
and in section 7, the effects of the torsion fields 
on the spin precession of a test body  are briefly analyzed
in view of an eventual experimental detection of the non-Riemannian  
contributions. 
\section{Equivalence of first and second order  formalisms}

Let us first give a short  review of 
the basic concepts of Riemann-Cartan geometry and
fix our notations and conventions.  
 For a complete introduction 
into the subject, the reader may consult Refs. \cite{1} and \cite{2}. 

Latin letters from 
the beginning of the alphabet ($a,b,c\dots $) run from 0 to 3 and 
are (flat) tangent space indices. Especially, $\eta_{ab}$ is the 
Minkowski metric $diag(1,-1,-1,-1)$ in tangent space. Latin letters 
from the middle of the alphabet ($i,j,k \dots $) are indices in a curved 
spacetime with metric $g_{ik}$ as before. We introduce the Poincar\'e gauge 
fields, the tetrad  $e^a_m $ and the connection 
$\Gamma^{ab}_{\ \ m}$ (antisymmetric in $ab$), as well as 
the corresponding field 
strengths, the curvature and torsion tensors
\begin{eqnarray}
R^{ab}_{\ \ lm} &=& \Gamma^{ab}_{\ \ m,l} - \Gamma^{ab}_{\ \ l,m} 
                     + \Gamma^a_{\ cl}\Gamma^{cb}_{\ \ m}   
- \Gamma^a_{\ cm}\Gamma^{cb}_{\ \ l}  \\ 
T^a_{\ lm} &=& e^a_{m,l} - e^a_{l,m} + e^b_m \Gamma^a_{\ bl}
- e^b_l \Gamma^a_{\ bm}. 
\end{eqnarray}
The spacetime connection $\Gamma^i_{ml} $ and the spacetime metric $g_{ik}$
can now be defined through 
\begin{eqnarray}
e^a_{m,l} + \Gamma^a_{\ bl} e^b_m &=& e^a_i \Gamma^i_{ml}\\
e^a_ie^b_k \eta_{ab} &=& g_{ik}.  
\end{eqnarray}
It is understood that there exists an inverse to the tetrad, such that 
$e^a_i e_b^i = \delta^a_b $. It can easily be shown that the connection 
splits into two parts, 
\begin{equation}
\Gamma^{ab}_{\ \ m} = \hat \Gamma^{ab}_{\ \ m} +
K^{ab}_{\ \ m},
\end{equation} 
such that $\hat \Gamma^{ab}_{\ \ m}$ is torsion-free 
and is essentially a function of $e^a_m$. $K^{ab}_{\ \ m}$ is 
 the contortion tensor (see below). Especially, 
the spacetime connection $\hat \Gamma^i_{ml}$ constructed from 
\begin{equation}
e^a_{m,l} + \hat \Gamma^a_{\ bl} e^b_m = e^a_i \hat \Gamma^i_{ml}
\end{equation}
is just the (symmetric) 
Christoffel connection of general relativity, a function of 
the metric only. 

The gauge fields $e^a_m $ and $\Gamma^{ab}_{\ \ m} $ are 
covector fields with respect to the spacetime index $m$. Under a local 
Lorentz transformation (more precisely, the Lorentz part of a 
Poincar\'e transformation, see Ref. \cite{3}) 
in tangent space, $\Lambda^a_{\ b}(x^m)$, 
they transform as 
\begin{equation}
e^a_m \rightarrow \Lambda^a_{\ b}e^b_m, \ \ \ \Gamma^a_{\ bm} \rightarrow 
\Lambda^a_{\ c}\Lambda^{\ d}_{b} \Gamma^c_{\ dm} - 
\Lambda^a_{\ c,m} \Lambda^{\ c}_{b}. 
\end{equation}
The torsion and curvature are Lorentz tensors with respect to their 
tangent space indices as is easily shown. The 
contortion $K^{ab}_{\ \ m}$ is also a Lorentz tensor and is related to 
the torsion through $K^i_{\ lm}= \frac{1}{2}(T_{l\ m}^{\ i}+ T_{m\ l}^{\ i}
- T^i_{\ \ lm})$, 
with $K^i_{\ lm} = e^i_a e^{}_{lb} K^{ab}_{\ \ m}$ and analogously 
for $T^i_{ \ lm}$. The inverse relation is $T^i_{\ lm} = 
-2 K^i_{\ [lm]}$. 

All quantities constructed from the torsion-free connection 
$\hat \Gamma^{ab}_{\ \  m}$ or $\hat \Gamma^i_{lm}$ will be denoted with 
a hat, for instance $\hat R^{il}_{\ \ km}
= e_a^i e_b^l \hat R^{ab}_{\ \ km}$ is the usual Riemann curvature tensor. 

We are now ready to compare the field equations arising in 
first and second order formalism and to establish their equivalence. 

Let $\mathcal L = \mathcal L_0 + \mathcal L_m$ be the Lagrangian 
density in the first order formalism and $\tilde {\mathcal L} = \tilde 
{\mathcal L}_0 + \tilde {\mathcal L}_m $ the same Lagrangian in the 
second order formalism. The indices $0$ and $m$ refer to the 
gravitational (or free) part and to the matter part respectively. 
Thus, we can write
\begin{equation}
\mathcal L(e^a_m, \Gamma^{ab}_{\ \ m}) 
= \tilde {\mathcal L}(e^a_m, K^{ab}_{\ \  m}).
\end{equation}
In concrete cases, the Lagrangians may differ by a total divergence, 
which does not influence the field equations and will therefore not be 
written out explicitely. For instance, in Ref. \cite{3}, we considered 
the teleparallel Lagrangian 
\begin{equation}
 e( R - \frac{1}{4}T^{ikl}T_{ikl}-\frac{1}{2}T^{ikl}T_{lki} 
+ \frac{1}{2}T^k_{\ ik}T^{mi}_{\ \ m}) = e \hat R,    
\end{equation}
where the left hand side is the first order form, while the right hand 
side is the usual Einstein-Hilbert Lagrangian expressed in terms of 
the tetrad only. In (11), a divergence term has been suppressed.
(In other words, the left and right hand sides are not really equal, 
but can be rendered equal by adding a physically irrelevant divergence.)

The field equations in the conventional,  
first order formalism are well known, 
\begin{equation}
\frac{\delta \mathcal L}{\delta \Gamma^{ab}_{\ \ m}} = 0\ \ \  \text{and} 
\ \ \ \frac{\delta \mathcal L}{\delta e^m_a} = 0. 
\end{equation}
As usual, they will be referred to as Cartan and Einstein equations,  
respectively. 
Both formalisms are related by equation (7), 
\begin{displaymath}
\Gamma^{ab}_{\ \ m} = \hat \Gamma^{ab}_{\ \ m}(e^a_m) + K^{ab}_{\ \ m}. 
\end{displaymath}
For the Cartan  equation in the second order formalism, we find 
\begin{equation}
\frac{\delta \tilde {\mathcal L}}{\delta K^{ab}_{\ \ m}} = 
\frac{\delta \mathcal L}{\delta \Gamma^{cd}_{\ \ l}}\  \frac{\delta
  \Gamma^{cd}_{\ \ l}}{\delta K^{ab}_{\ \ m}} = \frac{\delta \mathcal L}
{\delta \Gamma^{ab}_{\ \ m}} = 0. 
\end{equation}
Therefore, the Cartan equation has exactly the same form in both 
approaches. As to the Einstein  equation, we easily find 
\begin{equation}
\frac{\delta \tilde {\mathcal L}}{\delta e^m_a} 
= \frac{\delta \mathcal L}{\delta e^m_a} 
+ \frac{\delta \mathcal L}{\delta \Gamma^{cd}_{\ \ l}}\ \frac{\delta
  \Gamma^{cd}_{\ \ l}}{\delta e^m_a}
= \frac{\delta \mathcal L}{\delta e^m_a} +
\frac{\delta \mathcal L}{\delta \Gamma^{cd}_{\ \ l}}\ \frac{\delta
  \hat \Gamma^{cd}_{\ \ l}}{\delta e^m_a}=0.  
\end{equation}
This differs by the second term from the corresponding equation in (12).
This can be seen explicitely by 
 carrying  out the variation of this  term using the explicit form 
of $\hat \Gamma^{cd}_{\ \ l}$ from (8) and of the Christoffel symbols 
$\hat\Gamma^l_{ik}$. 

However, we can instead use immediately 
the Cartan equation $\delta \mathcal L / \delta \Gamma^{cd}_{\ \ l} = 0$ 
(which we have shown to be equivalent to the Cartan equation in the 
second order formalism) 
to eliminate the second term in (14), and therefore we 
have the on-shell relation $\delta \tilde {\mathcal L} / \delta e^m_a =  
\delta \mathcal L / \delta e^m_a $. 

To summarize, we have the following relations
\begin{equation} 
\frac{\delta \tilde {\mathcal L} }{ \delta e^m_a} = \frac{ 
\delta \mathcal L}{ \delta e^m_a }\ \ \ \text{and}\ \ \  
\frac{\delta \tilde {\mathcal L}}{\delta K^{ab}_{\ \ m}}  
= \frac{\delta \mathcal L}
{\delta \Gamma^{ab}_{\ \ m}}, 
\end{equation}
which shows clearly the equivalence of both sets of field equations. 
The above derivation also explains why some Lagrangians are allowed 
only in the first order formalism while others can only be considered 
in the second order formalism. The reason is that the terms 
resulting from the higher order derivatives (when 
 the \textit{wrong} formalism is applied to a certain Lagrangian), can 
actually be eliminated using the second field equation, thereby 
bringing the field equations into a form that would have resulted 
immediately if the \textit{right} formalism had been used right from 
the start.  (By \textit{wrong} and \textit{right}, we simply 
mean the formalism where we have higher order derivatives, or we have 
not, respectively. Both formalisms are correct, since they are equivalent.)

Note that in order to show the equivalence of both sets of field equations, 
we have to start with the full Lagrangian. We can of course not conclude 
that for given stress-energy tensor and spin density, the left hand 
sides of the field equations will have the same form in both procedures. 
The source tensors have to be modified at the same time.

Let us define the conventional canonical stress-energy and 
spin density tensors 
in the first order formalism 
\begin{equation}
T^a_{\ i} = \frac{1}{2e} \frac{\delta \mathcal L_m}{\delta e^i_a}\ \ \
\text{and}\ \ \ \sigma_{ab}^{\ \ m} = \frac{1}{e} \frac{\delta \mathcal L_m}
{\delta  \Gamma^{ab}_{\ \ m}}, 
\end{equation}
as well as their  second order analogues  
\begin{equation}
\tilde T^a_{\ i} = \frac{1}{2e} \frac{\delta \tilde{ \mathcal L}_m}
{\delta e^i_a}\ \ \
\text{and}\ \ \ \tilde \sigma_{ab}^{\ \ m} = \frac{1}{e} \frac{\delta 
\tilde 
{\mathcal  L}_m}{\delta  K^{ab}_{\ \ m}}.  
\end{equation}
Just as before (see (13) and (14)), one shows that 
\begin{equation}
\tilde \sigma_{ab}^{\ \ m} = \sigma_{ab}^{\ \ m}
\ \ \ \text{and}\ \ \ \tilde T^a_{\ i} = T^a_{\ i} + 
\frac{1}{2}\sigma_{cd}^{\ \  m}\ 
\frac{\delta \hat \Gamma^{cd}_{\ \ m}}{\delta e^i_a}. 
\end{equation}
Now, use the relation (8), $e^a_{i,k} + 
\hat \Gamma^a_{\ bk}e^b_i = e^a_l \hat \Gamma^l_{ik}$, and the explicit 
form of the Christoffel symbols $\hat \Gamma^l_{ik}$ to derive 
\begin{displaymath}
 \hat \Gamma^{cd}_{\ \ m} \!=\! \frac{1}{2}\left [
e^{ci}e^d_{i,m} \!-\! e^{di}e^c_{i,m}\!+\! e^{di}e^c_{m,i} 
\!  -\!e^{ci}e^d_{m,i}
\!+\! e^{di}e^{ck}e_{bm}e^b_{k,i}\!-\!  
e^{ci}e^{dk}e_{bm}e^b_{k,i}\right ]. 
\end{displaymath}
We can now evaluate $\tilde T^a_{\ i}$ in (18). The result is 
\begin{equation}
\tilde T^{lm} = T^{lm} +\frac{1}{2}
(\sigma^{lmk}- \sigma^{kml} - \sigma^{klm})_{;k},  
\end{equation}
  where $;$ denotes the usual covariant derivative with the Christoffel 
connection. 

Note that (19) looks basically like the Belinfante-Rosenfeldt 
relation between the 
symmetric Hilbert (or metric) stress-energy tensor $\tilde T^{lm}$ 
and the canonical (or Noether) tensor $T^{lm}$ known from purely 
Riemannian  theory, see Refs. \cite{1} and \cite{2}, especially the 
discussion on relocalizations in Ref. \cite{2}). We will see  
that $\tilde T^{lm}$ is not yet the symmetric tensor, but it 
is a tensor that reduces to the symmetric Hilbert tensor 
for vanishing torsion. 

In order to prevent any kind of misunderstandings, let us emphasize 
once again that we do not modify the underlying gauge structure of 
the theory. Especially, the relation (10) makes clear the we do 
not change the coupling prescriptions in the matter Lagrangians. Thus, 
for instance, the Dirac particle still couples minimally to the 
gauge connection $\Gamma^{ab}_{\ \ i}$ (which, from a mathematical 
point of view, is the fundamental field variable). The only difference 
is that we express this connection in terms of  $e^a_i $ and  
$K^{ab}_{\ \ i}$. Any change in the coupling prescription would lead to 
a fundamentally  different theory, especially concerning  the gauge 
symmetry.

In the next section, we will derive the conservation equations for 
spin density and stress energy tensor in both approaches.

\section{The Noether identities}
\subsection{First order formalism}
Let us briefly review the derivation of the Noether identities 
related to local, tangent space Lorentz transformations and  to general 
coordinate transformations. Under an infinitesimal Lorentz transformation (9),
with $\Lambda^a_{\ b} = \delta^a_b + \epsilon^a_{\ b}$ 
($\epsilon^{ab}=-\epsilon^{ba}$) 
the independent fields $e_a^m$ and $\Gamma^{ab}_{\ \ m}$ undergo 
the following change:
\begin{equation}
\delta \Gamma^{ab}_{\ \ m} = - \epsilon^{ab}_{\ \ ,m}- \Gamma^a_{\ cm}
\epsilon^{cb} - \Gamma^b_{\ cm} \epsilon^{ac}\ \ \ \text{and}\ \ \  
\delta e^m_a = \epsilon_a^{\ c} e^m_c. 
\end{equation}
The first equation can be written in the short form $\delta \Gamma^{ab}_{\ \
  m}  = - \De_m 
\epsilon^{ab}$. 
The change in the matter Lagrangian therefore reads
\begin{equation}
\delta \mathcal L_m = \frac{\delta \mathcal L_m}
{\delta e^m_a} \delta e^m_a 
+ \frac{\delta \mathcal L_m}{\delta \Gamma^{ab}_{\ \ m}}
\delta \Gamma^{ab}_{\ \ m}= e(2T^{[ac]}+ \De_m \sigma^{acm})\epsilon_{ac},  
 \end{equation}
where we have omitted a total divergence. The covariant 
derivative operator $\De_m$ is defined to act with $\Gamma^{ab}_{\ \ m}$ 
on tangent space indices and with $\hat \Gamma^l_{ki}$ (torsion free) 
on spacetime indices. The requirement of Lorentz gauge invariance 
therefore leads to 
\begin{equation} 
2T^{[ac]}+ \De_m \sigma^{acm}= 0. 
\end{equation}
Slightly more complicated is the case of the coordinate invariance. 
Under an infinitesimal coordinate transformation, 
\begin{equation}
\tilde x^i = x^i + \xi^i, 
\end{equation}
the fields $\Gamma^{ab}_{\ \ m}$ and $e^a_m$ transform as spacetime covectors
(or one-forms), while the inverse tetrad $e^m_a$ is a contravariant vector 
with respect to the index $m$. Thus, 
\begin{displaymath}
\tilde e^m_a(\tilde x) = e^m_a(x) + \xi^m_{\ ,k} e^k_a 
\ \ \ \text{and}\ \ \ \tilde \Gamma^{ab}_{\ \ m}(\tilde x) = 
\Gamma^{ab}_{\ \  m}(x) - \xi^k_{\ ,m}\Gamma^{ab}_{\ \ k}. 
\end{displaymath}
Since we are interested in the change of the Lagrangian under an 
active transformation, we have to  evaluate the change of the fields  
at the same point $x$, and thus  have to express the transformed fields 
in the old coordinates, i.e., 
\begin{displaymath}
\tilde e^m_a(x) = \tilde e^m_a(\tilde x) - \tilde e^m_{a,k}(\tilde x) 
\xi^k,\ \ 
\tilde \Gamma^{ab}_{\ \ m}(x) = \tilde \Gamma^{ab}_{\ \ m} (\tilde x)-
\xi^k \tilde \Gamma^{ab}_{\ \ m,k}(\tilde x). 
\end{displaymath}
In the $\xi$-terms, we can replace $\tilde e^m_a(\tilde x)$ by $e^m_a(x)$ and  
$\tilde\Gamma^{ab}_{\ \ m}(\tilde x)$ by $\Gamma^{ab}_{\ \ m}(x)$, since the 
difference will be of order $\xi^2$. 
Finally, we find
\begin{eqnarray}
\delta e^m_a &=&
 \tilde e^m_a(x) - e^m_a(x) = \xi^m_{\ ,k}e^k_a- \xi^k e^m_{a,k}, \\
\delta \Gamma^{ab}_{\ \ m} &=& \tilde \Gamma^{ab}_{\ \ m}(x) - 
\Gamma^{ab}_{\ \ m}(x) = - \xi^k_{\ ,m}\Gamma^{ab}_{\ \ k}- \xi^k
 \Gamma^{ab}_{\ \ m,k}.  
\end{eqnarray}
Let us note at this point that, if we define the parameters 
$\epsilon^a = - e^a_m \xi^m $ and $\epsilon^{ab} = \xi^k \Gamma^{ab}_{\ \ k}$, 
the above transformations can be written in the form 
\begin{eqnarray*}
\delta e^a_m &=& \epsilon^a_{\ c}e^c_m - \De_m \epsilon^a - \epsilon^l T^a_{\
  lm}\\  
\delta \Gamma^{ab}_{\ \ m} 
&=& - \De_m \epsilon^{ab} - \epsilon^l R^{ab}_{\ \  lm}. 
\end{eqnarray*}
Since we are also free to perform an additional, independent,  
Lorentz transformation (20), we 
may again consider $\epsilon^a$ and $\epsilon^{ab}$ as independent of each 
other. 
In this form, the transformations are sometimes referred to as 
generalized Poincar\'e gauge transformations, with $\epsilon^a$ generating the 
translations and $\epsilon^{ab}$ the Lorentz rotations \cite{1}. 
We will remain in the usual interpretation of active coordinate 
transformations and return to the form (24)-(25). 
The change of the Lagrangian reads 
\begin{eqnarray*}
  \delta \mathcal L_m &=& \frac{\delta L_m}{\delta e^m_a} \delta e^m_a 
+ \frac{\delta \mathcal L_m}{\delta \Gamma^{ab}_{\ \ m}} \delta 
\Gamma^{ab}_{\  \ m} \\  &=&\! \! \! \!   
-2(eT^a_{\ m}e^k_a)_{,k}\xi^m \!\! -\! 2eT^a_{\ m}e^m_{a,k}\xi^k
\!\!+\! (e\sigma_{ab}^{\ \ m}\Gamma^{ab}_{\ \ k})_{,m} \xi^k 
\!\! -\! e\sigma_{ab}^{\ \ m}
\Gamma^{ab}_{\ \ m,k} \xi^k. 
\end{eqnarray*}
Requiring $\delta \mathcal L_m = 0$ and regrouping 
carefully the terms, we finally 
get 
\begin{equation}
(\De_m T^m_{\ \ b}) e^b_k + T^m_{\ \ b} T^b_{\ mk} 
= \frac{1}{2}R^{ab}_{\ \ mk} 
\sigma_{ab}^{\ \ m} + \frac{1}{2}\Gamma^{ab}_{\ \ k} (\De_m\sigma_{ab}^{\ \ m}
+ 2 T_{[ab]}). 
\end{equation}
The last term vanishes with (22) and the first two terms can be 
rewritten in a different form, using the relation  $T^i_{\ lm}=
-2 K^i_{\ [lm]}$ in the second term, and  the explicit form of the 
covariant derivative $\De_m$ in the first term, 
to get 
\begin{equation}
T^{m}_{\ k;m} - K^{im}_{\ \ k} T_{mi} = \frac{1}{2}R^{ab}_{\ \
  mk}\sigma_{ab}^{\ \ m}. 
\end{equation}
Recall that $K^{im}_{\ \ k}$ is antisymmetric in $im$, so that 
classical matter (with $T_{[mi]}= 0$ and $\sigma_{ab}^{\ \ m} = 0$) 
will satisfy the general relativistic conservation law $T^{km}_{\ \ ;m} = 0$. 

\subsection{Second order formalism}

In the second order formalism, we have to evaluate the changes of 
the fields $e^m_a$ and $K^{ab}_{\ \ m}$ under Lorentz and 
coordinate transformations. The Lorentz transformation now reads 
\begin{equation}
\delta e^m_a = \epsilon_a^{\ c} e^m_c, \ \ \ \delta K^{ab}_{\ \ m}= 
\epsilon^a_{\ c} K^{cb}_{\ \ m} + \epsilon^b_{\ c} K^{ac}_{\ \ m}, 
\end{equation}
which shows clearly that $K^{ab}_{\ \ m}$ transforms as Lorentz tensor. 
Evaluating 
\begin{displaymath}
\delta \tilde {\mathcal L}_m = \frac{\delta \tilde {\mathcal L}_m}
{\delta e^m_a} 
\delta e^m_a + \frac{\delta \tilde {\mathcal L_m}}{\delta K^{ab}_{\ \ m}} 
\delta K^{ab}_{\ \ m}, 
\end{displaymath}
and requiring $\delta \tilde {\mathcal L}_m = 0$ 
we easily derive the following 
relation
\begin{equation}
0 = 2\tilde T^{[ac]} - K^{bc}_{\ \ m} \sigma^{a\ m}_{\ b} + K^{ba}_{\ m}
\sigma^{c\ m}_{\ b}. 
\end{equation}
This does not look like a conservation equation for the spin density. 
We see that in the case where the torsion vanishes, the stress-energy 
tensor becomes symmetric. Therefore, we can see $\tilde T^{im}$ as 
a generalization of the symmetric Hilbert tensor. Let us note 
at this point that if we had considered as independent the fields 
$K^{ik}_{\ \ l}$ and $e^a_m$ (instead of $K^{ab}_{\ \ l}$ and $e^a_m$), 
we would have found a stress-energy tensor that is always symmetric. 
This, however, does not present any advantages (the variation of the 
gravitational Lagrangian would in most cases become more complicated) 
and so we preferred to stay a step closer to the conventional 
gauge approach, considering $K^{ab}_{\ \ m}$ with its natural 
tangent space indices. The advantage of this approach is the 
linear relation between $\Gamma^{ab}_{\ \ m}$ and $K^{ab}_{\ \ m}$ and 
 the resulting complete equivalence of the Cartan field equation in first 
and second order formalism. 
The alternative approach,  with $K^i_{\ lm}$ and 
$e^a_m$,  has been used in the framework of Einstein-Cartan-Dirac theory
in the past \cite{4}. It leads of course to a symmetric Einstein equation. 

Under coordinate transformations, the fields transform exactly as 
in the previous section (because just as $\Gamma^{ab}_{\ \ m}$,  
$K^{ab}_{\ \ m}$ too is a spacetime covector), i.e., 
\begin{eqnarray}
\delta e^m_a &=&
 \tilde e^m_a(x) - e^m_a(x) = \xi^m_{\ ,k}e^k_a- \xi^k e^m_{a,k}, \\
\delta K^{ab}_{\ \ m} &=& \tilde K^{ab}_{\ \ m}(x) - 
K^{ab}_{\ \ m}(x) = - \xi^k_{\ ,m}K^{ab}_{\ \ k}- \xi^k
K^{ab}_{\ \ m,k}.  
\end{eqnarray}
In the same way as in the previous section, we derive the conservation 
equation in the form 
\begin{eqnarray*}
  (\De_m \tilde T^m_{\ a})e^a_k + \tilde T^m_a T^a_{\ mk} - \frac{1}{2} 
\sigma_{li\ \ ;m}^{\ \ m} K^{li}_{\ \ k}  \\             
 = \frac{1}{2} \left[ 
K^{ab}_{\ \ k,m}- K^{ab}_{\ \ m,k} 
+ \hat \Gamma^a_{\ cm} K^{cb}_{\ \ k} 
+ \hat \Gamma^b_{\ cm}K^{ac}_{\ \ k}  
 \right. \\ \left. - \hat \Gamma^a_{\ ck} K^{cb}_{\ \ m} 
- \hat \Gamma^b_{\ ck}K^{ac}_{\ \ m} 
+ K^a_{\ cm}K^{cb}_{\ \ k} + K^b_{\ cm}K^{ac}_{\ \ k}\! \right ] 
\sigma_{ab}^{\ \ m}.
\end{eqnarray*}
We used (29) to put the equation into this form. 
A shorter way of writing  this equation is the following 
\begin{equation}
(\De_m \tilde T^m_{\ \ a})e^a_k + \tilde T^m_{\ a} T^a_{\ mk} - \frac{1}{2} 
\sigma_{li\ \ ;m}^{\ \ m} K^{li}_{\ \ k} = 
\frac{1}{2}(R^{ab}_{\ \ mk}- \hat R^{ab}_{\ \ mk}) 
\sigma_{ab}^{\ \ m}. 
\end{equation}

\subsection{Discussion}

Comparing (22) and (29), we can derive the following relation 
between the antisymmetric parts of the stress-energy tensors:
\begin{equation}
2(\tilde T^{[lm]} - T^{[lm]} ) = \sigma^{lmk}_{\ \ \ ;k}.  
\end{equation}
This relation is easily verified using the explicit form of 
$\tilde T^{lm}$ in terms of $T^{lm}$, Eq. (19). 

Also, from (26) and (32), we get a relation between the divergences 
of the stress-energy tensors of the form 
\begin{equation}
(T^m_{\ k} - \tilde T^m_{\ k})_{;m} = \frac{1}{2} \hat R^{ab}_{\ \ mk}
\sigma_{ab}^{\ \ m}. 
\end{equation}
We used (33) to bring the relation into this form. Again, this relation 
can be derived directly from (19), using the symmetry properties 
of $\sigma^{lmk}$ and of the Riemann tensor $\hat R^{i}_{\  lkm}$. 

Note that both sets of equations, (22,26) and (29,32),  are 
valid, it is not the one \textit{or} the other set. (The fact that the 
Lagrangian is invariant under Lorentz gauge and under 
general coordinate transformations does not depend on the choice of 
the independent field variables.) 
They are simply different 
equations for different quantities. The important thing is to 
use the right quantities as right hand side of the 
gravitational field equations (this depends on the choice of our 
formalism, first or second order).

Another point concerns the physical interpretation of $T^{ik}$ and 
$\tilde T^{ik}$. If we take the point of view that the measurable 
quantities are not the densities, but rather the integrated quantities, 
namely the momentum vector and the spin tensor \cite{5}, then 
it is easy to see that $T^{ik}$ and $\tilde T^{ik}$ lead, in general,
to different definitions of the momentum. Apart from  this ambiguity, 
there is the problem that even from a single stress-energy tensor, 
one can define many inequivalent momentum vectors, integrating 
$T^a_{\ i}$, $T^i_{\ a}$, $T^{ik}$ etc. (see Ref. \cite{6} 
for an interesting 
discussion on this point). 
Usually, one would like to define the momentum in a way that  
is conserved. This is of course not possible in our case, since 
whatever momentum vector one defines, it will always be subject to 
gravitational forces coming from curvature and torsion, and therefore 
it cannot be conserved. (The other way around, whatever momentum vector 
one uses, it will always be conserved in the limit of vanishing 
curvature and torsion.)
As it seems, it is rather a matter of convention 
how one wishes to define spin and momentum, under the only restriction 
 that for vanishing  fields, they should be conserved.   

In view of the rather unusual equation (29), which does not look like 
a conservation equation, one might be tempted to reject the set of 
current densities $\tilde T^a_{\ m}, \sigma_{ab}^{\ \ m}$ and to claim 
that the canonical currents $T^a_{\ m}, \sigma_{ab}^{\ \ m}$ as they 
arise in the conventional, first order formalism, should be considered as 
the true, physical currents. However, let us recall  the fact that, 
for instance, the canonical stress-energy tensor $T^a_{\ m}$ of the 
Dirac Lagrangian contains non-axial torsion parts, which, as is well 
known, do not couple to the Dirac particle in the first place,  
and are neither contained 
in the Dirac equation, nor in the Lagrangian (where they cancel out
with the hermitian conjugated term). We have brought this to attention 
in Ref. \cite{5}, to show that one has to be very careful on how to extract 
physical quantities from $T^{a}_{\ m}$. Thus, neither can the 
conventional, first order canonical tensors, be used 
without precautions to derive physically valid results.

In order to gain at least some insight into the contents of $T^{ik}$ and
 $\tilde T^{ik}$, let us take a look at the flat limit of our 
equations, i.e.,  assume a 
 spacetime with $g_{ik} = \eta_{ik}$. 
In this case, we find from (19) 
\begin{equation}
\tilde T^{lm} = T^{lm} +\frac{1}{2}
(\sigma^{lmk}- \sigma^{kml} - \sigma^{klm})_{,k}.   
\end{equation}
If we define the momentum vector as we do in special relativity, 
\begin{equation}
P_k = \int T^i_{\ k}\ \de S_i,
\end{equation}
then, for the alternative stress-energy tensor, we have 
\begin{equation}
\tilde P_k = \int \tilde T^i_{\ k}\ \de S_i = P_k + \frac{1}{2} \int 
(\sigma^{i\ \ l}_{\ m}- \sigma^{l\ \ i}_{\ m} - 
\sigma^{li\ }_{\ \ m})_{,l}\  \de S_i. 
\end{equation}
Clearly, the last term can be converted into a 4d volume integral over 
the divergence of the integrand, which vanishes however, due to 
the antisymmetry in $il$ of the expression in brackets. Thus, we have 
\begin{equation}
\tilde P_k = P_k. 
\end{equation}
Of course, we also have (see (34) or (35)) 
\begin{equation}
\tilde T^m_{\ \ k,m} = T^m_{\ \ k,m}. 
\end{equation} 
Usually, the relations (38) and (39) are said to show the equivalence of the 
two stress-energy tensors. (Same charges, same conservation law.) 
More generally, a change of the stress-energy tensor and the spin in the form 
\begin{eqnarray}
T^{lm} & \rightarrow & T^{lm} + X^{mlk}_{\ \ \ ,k}, \\
\sigma_{km}^{\ \ i} & \rightarrow & \sigma_{km}^{\ \ i} - (X_{km}^{\ \ i} 
- X_{mk}^{\ \ i}) 
\end{eqnarray}
with $ X^{mkl} = - X^{mlk}$ 
is called relocalization of stress-energy and spin and is said not to 
affect physical quantities. 

It is easily seen that (35) is a relocalisation of $T^{lm}$, without, 
however, the corresponding transformation of the spin density.  
 Nevertheless, (26) and (29) coincide in flat 
space. The remaining relation is
\begin{eqnarray*}
 \tilde T^m_{\ k,m} &=& T^m_{\ k,m} \\  
 &=& \frac{1}{2}K^{im}_{\ \ k}  
 (\sigma_{im\ ,l}^{\ \ l} - K^j_{\ il}\sigma_{jm}^{\ \ l}
- K^j_{\ ml}\sigma_{ji}^{\ \ l}) 
 \\
&&+ \frac{1}{2}
(K^{ab}_{\ \ k,m}-K^{ab}_{\ \ m,k}+K^a_{\ cm}K^{cb}_{\ \ k}
+K^a_{\ cm}K^{ac}_{\ \ k}) \sigma_{ab}^{\ \ k}.  
\end{eqnarray*}

Unfortunately, we cannot express the equation for the spin precession 
(22) or (29) in a form independent of the  stress-energy tensor. 
(Using (35) in (33) leads to an identity.)
However, since the spin density 
is the same in both approaches, it will be subject to the 
 same equations of 
motion, expressed in one or the other way. 

We therefore conclude that in the flat limit, $g_{ik} = \eta_{ik} $,
the equations of motions for the current densities 
obtained in both pictures are identical 
as far as their physical contents is concerned. 

In curved spacetime, this is not the case anymore. Especially, we get 
two different equations for the stress-energy tensor and we have to 
decide, based on physical arguments, how to define  define 
the momentum vector. 

Let us once again refer the reader to the extended discussion on 
canonical and metric stress-energy tensors and their relation via 
relocalizations in 
the general framework of metric affine theories in Hehl et al., \cite{2}. 

\section{Model construction}

Comparing  equations (27) and (32), we see that the following tensor 
\begin{eqnarray}
F^{ab}_{\ \ mk}&=&  
\frac{1}{2}(R^{ab}_{\ \ mk}- \hat R^{ab}_{\ \ mk}) \nonumber \\
&=&K^{ab}_{\ \ k,m}- K^{ab}_{\ \ m,k} 
+ \hat \Gamma^a_{\ cm} K^{cb}_{\ \ k} 
+ \hat \Gamma^b_{\ cm}K^{ac}_{\ \ k} \nonumber \\ 
&&- \hat \Gamma^a_{\ ck} K^{cb}_{\ \ m} 
- \hat \Gamma^b_{\ ck}K^{ac}_{\ \ m} 
+ K^a_{\ cm}K^{cb}_{\ \ k} + K^b_{\ cm}K^{ac}_{\ \ k} 
\end{eqnarray}
plays, in the context of the second order formalism, a similar role 
as the curvature tensor $R^{ab}_{\ \ lm}$ in the conventional approach. 
Essentially, $F^{ab}_{\ \ mk}$ has the structure of a Lorentz Yang-Mills 
tensor, but with $\Gamma^{ab}_{\ \ m}$ replaced by $K^{ab}_{\ \ m}$. The 
terms containing the Christoffel connection $\hat \Gamma^{ab}_{\ \ m}$ 
are needed to get the correct tensor behavior under a 
Lorentz transformation. 

Apart from $F^{ab}_{\ \ mk}$, which we will use to produce the dynamical, 
Yang-Mills like, behavior of the torsion field, we will need  
a term that guarantees the correct general relativity limit of the theory. 
In the second order formalism, this is simply the Einstein-Hilbert Lagrangian, 
expressed in terms of the tetrad field, $\hat R$. For reasons that will 
become clear later, we also include a mass term $K^{ab}_{\ \ m}K_{ab}^{\ \ m}$ 
into our Lagrangian. 

Thus, we start with 
\begin{equation}
\mathcal L_0 = e(a \hat R - b F^{ab}_{\ \ mk}F_{ab}^{\ \ mk} - c K^{ab}_{\ \ m}
K_{ab}^{\ \ m}). 
\end{equation}
This Lagrangian contains no second order derivatives of the independent 
fields $e^a_i$ and $K^{ab}_{\ \ i}$, apart from the well known 
higher order terms contained in $\hat R$, which can, however, be 
eliminated by subtracting a surface term. On the other hand, 
if $\mathcal L_0$ were to be treated in the first order formalism, 
there would arise higher order derivatives in the tetrad fields, coming 
from the term $\hat R^{ab}_{\ \ lm}\hat R_{ab}^{\ \ lm}$ contained in 
the second term of (43).

Including the matter Lagrangian $\mathcal L_m$, we are let to the 
following set of field equations
\begin{eqnarray}
-4b\  \De_k F_{ab}^{\ \ mk} +2c K_{ab}^{\ \ m} &=& \sigma_{ab}^{\ \ m} \\
\ \ \ \ \ \ \ \ \ \ \ \ \ \  \ \ \ \ \ \ \  
-a \hat G_{ki} &=& \tau^{(1)}_{ ki} + \tau^{(2)}_{\ ki}
+ \tau^{(3)}_{\ ki} + \tilde T_{ki}, 
\end{eqnarray}
where 
\begin{eqnarray}
\tau^{(1)}_{\ ki}&=& - 2b(F^{cb}_{\ \ mi}F_{cb\ \ k}^{\ \ m}-\frac{1}{4}g_{ki} 
F^{cb}_{\ \ ml}F_{cb}^{\ \ ml}) \\
\tau^{(2)}_{\ ki} &=&
- c(K^{cb}_{\ \ k}K_{cbi}-\frac{1}{2}g_{ki} K^{cb}_{\ \ l}K_{cb}^{\ \ l})\\
\tau^{(3)}_{ki} &=& 
-2b\left[ (F_{imk}^{\ \ \ n}K^{lm}_{\ \ n})_{;l}
+(F_{in}^{\ \ lm}K_{k\ m}^{\ n})_{;l}
-(F_{km}^{\ \ ln}K_{i\ n}^{\ m})_{;l} \right. \nonumber \\
&&\left. -(F^{l}_{\ mkn}K_{i}^{\ mn})_{;l}
+(F_{kmi}^{\ \ \ n}K^{lm}_{\ \ n})_{;l}
-(F^{l\ \ n}_{\ mi}K_{k\ n}^{\ m})_{;l}\right]. 
\end{eqnarray}
We recognize in (44) a Yang-Mills type equation for the field $K^{ab}_{\ \
  m}$, endowed with a mass term proportional to $c$. 
(Recall that we have defined $\De $ to act with the full Lorentz connection 
$\Gamma^{ab}_{\ \ m}$ on tangent space indices, and with the Christoffel 
symbol on spacetime indices.) The corresponding 
symmetric, traceless stress-energy tensor is found in (46) and 
the stress-energy contribution from the mass term is given by (47). 

The only non-symmetric contributions in (45) are contained in 
the unusual term 
$\tau^{(3)}_{\ ki}$, a term that results from the $\hat \Gamma^{ab}_{\ \ m}$ 
couplings in $F^{ab}_{\ \ mk}$. In this term, 
the Lorentz gauge structure of the theory is not obvious, since the spacetime 
and Lorentz indices of $F^{ab}_{\ \ mk}$ and $K^{ab}_{\ \ m}$ 
are completely mixed up. This is also the reason why 
we chose to write down Eq. (45) immediately in terms of spacetime tensors, 
as opposed to the original form $\hat G^a_{\ i} = \hat T^{a}_{\ i} + \dots $ 
that results from the variation with respect to $e^i_a$. In contrast 
to (44), which respects clearly the  underlying gauge structure, 
nothing useful is 
revealed by keeping (45) in its mixed form, and it is convenient to pull 
down all quantities to physical spacetime such that one has to deal 
only with one kind of indices.     

Concretely, we find for the antisymmetric part of  (45)  
 \begin{equation}
\tilde T^{[ik]} = - 2b (F_{\ n}^{i\ lm}K^{kn}_{\ \ m}-
F^{k\ lm}_{\ n}K^{in}_{\ \ m})_{;l}.
\end{equation}
The calculations are rather lengthy, but one can directly verify that 
(49), together with (44), leads to the Noether identity (29).    

In the classical, spinless limit, it is immediately clear that 
(44) leads to the groundstate solution $K^{ab}_{\ \ m} =0$, while 
(45) reduces to 
\begin{displaymath}
\hat G_{ik} =  T_{ik}, 
\end{displaymath} 
which are the field equations of general relativity. (We took the 
opportunity to fix the parameter $a$ to $a = -1$ to be consistent 
with the conventional choice of units.) Also, in view of (19), there is no 
need, in the classical limit, to distinguish between $T^{ik}$ 
and $\tilde T^{ik}$. 

Thus, the theory described by (43) allows for a complete general relativity 
limit in the absence of spinning matter, while a source with intrinsic 
spin, e.g., an elementary particle or a spin polarized neutron star, 
will give rise to dynamical torsion fields via (44). 
This cannot be achieved within the conventional first order formalism
 if one is not willing to allow for higher order derivatives,  
and therefore, no such theory has yet been presented.  
In order to avoid misunderstandings, let us emphasize that this does not mean 
that there are no consistent Poincar\'e gauge theories based on the 
first order formalism without higher derivatives. 
What it means is that such theories do not 
have a full general relativity limit. Therefore, constraints on 
the free parameters of the theory will arise from the classical 
experiments, based on planetary motion and light propagation. This 
is rather disappointing, since it essentially means generalizing   
GR, but  holding  the changes \textit{small} enough to ensure that they do 
not affect classical physics in a sensible way. On the other hand, 
in our model, classical measurements are not affected at all, independently 
of the coupling constants $b,c$, and therefore, the post-general 
relativistic  effects will only arise if spinning matter is introduced. 
Thus, the generalization affects  only  spinning  matter, in view 
of which the theory was generalized in the first place. In short: Additional 
fields (torsion) for additional degrees of freedom (spin). 

Finally, let us make a remark on the mass term in (43). The reason 
for its introduction is easily seen in context of the classical limit. 
Suppose $c=0$ in (44)-(45). Then, every GR solution, satisfying 
in addition  $F^{ab}_{\ \ mk}=0$, will be solution to the field 
equations with $\sigma_{ab}^{\ \ m}=0$. However, from $F^{ab}_{\ \ mk} =0$, 
we can not conclude to the unique solution $K^{ab}_{\ \ m}=0$ 
(we will see an explicit example in the next section). However, 
changing $K^{ab}_{\ \ m}$ by a term that leaves $F^{ab}_{\ \ mk}$ unaffected 
will have physical consequences (e.g., concerning the spin precession of a 
test particle) and therefore, this additional freedom cannot be accepted. 
This is not confined to the classical limit. From (43), it is obvious 
that, for $c=0$, the field equations will only determine $F^{ab}_{\ \ mk}$, 
but not $K^{ab}_{\ \ m}$, allowing therefore, in general, for a whole class 
of solutions for $K^{ab}_{\  \ m}$. This will become very clear in the 
context of the solutions we will analyze in the next section. 
The introduction of the mass term obviously resolves this problem. 

\section{Yasskin type reduction}

Since the classical limit of our theory coincides completely with 
general relativity, the only way to test the new features related to 
the presence of torsion fields is to look at field configurations 
generated by a source with intrinsic spin. One way is for instance 
to study interactions between elementary particles. On the astrophysical 
level, one can consider massive, spin polarized bodies, especially 
neutron stars, and analyze, e.g., the evolution of a system 
of two such bodies or, which is simpler from the computational point of 
view, study the motion and spin precession of a test body (a smaller 
neutron star or elementary particles) in the vicinity of a central 
source.  

In this section, we try to get solutions for the fields at the exterior of 
such a macroscopic spinning body. As discussed extensively in Ref. 
\cite{5}, 
such a matter distribution (located at the origin) is described by a 
spin density of the form 
\begin{equation}
\sigma_{ab}^{\ \ m} = \sigma_{ab} u^m \rho(x), 
\end{equation}
where $\rho(x)$ is a suitably normalized density function and 
$\sigma_{ab}$ is the integrated spin tensor. This spin density also 
coincides with that describing the generalized Weyssenhoff fluid 
in Riemann-Cartan spacetimes \cite{7}. Then, following Yasskin \cite{8}, 
we look for solutions that have essentially the same structure 
as the current density, but in view of our rather special Lagrangian, 
we make the ansatz not for the Lorentz gauge potential $\Gamma^{ab}_{\ \ m}$, 
but for the contortion $K^{ab}_{\ \ m}$, i.e., we set  
\begin{equation}
K^{ab}_{\ \ m} = \sigma^{ab} A_m, 
\end{equation}
where the field $A_m$ will have to be determined by the field equations. 
In Ref. \cite{8}, 
with a similar ansatz, the Einstein-Yang-Mills field equations 
were reduced essentially to an Einstein-Maxwell system of equations, under 
the assumption that the gauge charges (in our case, $\sigma_{ab}$) are 
constant. In order to preserve Lorentz invariance, one cannot require 
$\sigma_{ab}$ to be constant, but instead has to use the constraint  
\begin{equation}
\hat \De_i \sigma_{ab} = 0, 
\end{equation}
which, with the help of (51), can also be written in the form 
\begin{equation}
\De_i \sigma_{ab} = 0.  
\end{equation}
In other words, the contortion fields, which in our second order approach 
play a role similar to the Yang-Mills potentials, drop out from  the 
covariant derivative of $\sigma_{ab}$. This is 
similar to Ref. \cite{8}, where the gauge covariant derivatives coincide, 
for the specific ansatz, with the partial derivatives, which makes 
the requirement that the charges be constant a covariant one. 

What does Eq. (52) mean physically? Well, in view of a physical
interpretation of the spin tensor, one will have to use further 
constraints, the so-called spin supplementary condition (SSC), like 
$\sigma_{ik}u^i = 0$ or similar. Roughly, there will  be a local 
coordinate system where (52) reduces to $\sigma_{ik} = const$ and 
one might then choose the SSC $\sigma_{i0} = 0$ and define the 
spin vector $\sigma^{\mu} = \epsilon^{\mu \nu \lambda} \sigma_{\nu \lambda}$
($\mu, \lambda, \nu = 1,2,3$). 
In other words, (52) is simply the covariant expression of the fact 
that the source is described by a constant spin vector. (Constant 
in the sense of a fixed vector, pointing in a fixed direction, 
as opposed to a radially orientated, spherically symmetric spin vector 
field, for instance.) This is just what we expect for a spin polarized 
body. Note, on the occasion, that $\sigma_{ab}\sigma^{ab} 
= 2 \vec \sigma^2$. Therefore, let us define the spin $s$ through 
\begin{equation}
\sigma_{ab}\sigma^{ab} = 2 s^2.
\end{equation}

In practice, we will apply  (52) only to the contortion field (51), and not   
to (50), since we will only deal with exterior solutions here. The 
above arguments  are therefore only a simplified version of 
what should be the result of an analysis of the interior dynamics of  
the neutron star, which is beyond the scope of this article. For our 
purposes, the star can equally well be considered  to be pointlike 
(i.e.,  $\rho(x) = \delta(x)$ in (50)), as long as (52) is satisfied. 

With the ansatz  (50)-(52), the field tensor (42) reduces to 
\begin{equation}
F^{ab}_{\ \ ik} = \sigma^{ab} F_{ik}, 
\end{equation}
where  $F_{ik} = A_{k,i}- A_{i,k}$, whereas  the field equations (44) and (45) 
take the simple form 
\begin{eqnarray}
 - 4 b F^{mk}_{\ \ ;k} &=&  
- 2c A^m + \rho(x) u^m \\
\hat G_{ik} &
=&      - 4b s^2(F_{mi}F^m_{\ k}- 
\frac{1}{4} g_{ik} F^{lm}F_{lm}) \nonumber \\ 
&&- 2 c s^2 (A_kA_i - \frac{1}{2} g_{ik}A_l A^l) + \tilde T_{ik}.
\end{eqnarray}
Thus, with a slight extension of the method presented in Ref. \cite{8}, 
we have brought our equations into an Einstein-Proca system. 

Several remarks are in  order at this point. First, the ansatz (51) 
is supposed to hold at the exterior of the source. We do not know 
anything about the field configurations inside the source. Nevertheless, 
we have included the source terms in (56)-(57) in order to be able to 
fix eventual parameters of the solutions (like the Schwarzschild mass etc.). 
You may consider the source terms as boundary conditions, having in mind that 
the above form of the field equations is valid only at the exterior.  
Secondly, and not unrelated to the first remark, you may have noticed 
 that equation 
(57) is symmetric, while one  expects an asymmetric stress-energy 
tensor for a spinning body. 
 This, of course, reflects again the  fact that 
the equations do not hold inside the matter distribution. In order to 
ensure the continuity of the solutions at the boundary of the source, 
the stress-energy tensor should have a symmetric limit at that boundary. 
Recall however that the tensor $\tilde T_{ik}$, as we have outlined at  
the end of section 2, is not the canonical stress-energy tensor, but 
a tensor that is already quite close to   the symmetric Hilbert tensor. 
More precisely, we have shown that its 
antisymmetric part is given by (see (29))
\begin{displaymath}
 2\tilde T^{[ac]} = K^{bc}_{\ \ m} \sigma^{a\ m}_{\ b} - K^{ba}_{\ m}
\sigma^{c\ m}_{\ b}. 
\end{displaymath}
Using the spin density (50) and assuming that $K^{ab}_{\ \ m}$ tends 
to the form (51) as we approach the boundary, we see that 
$T^{[ac]}$ vanishes without any restrictions on the stress-energy tensor. 
Therefore, the symmetry of (57) is not a constraint on the stress-energy 
tensor, but simply the result of the specific form  (51) of the torsion.  

On the other hand, if we assume that $K^{ab}_{\ \ m}$ has a similar form  
also at the interior of the source (this is to be  understood as 
the mean value 
over a macroscopic region, not on an elementary particle level), 
then we can conclude that 
the stress-energy tensor is symmetric allover, and that it essentially 
coincides with the Hilbert tensor of general relativity. This is 
of course in contrast to conventional, first order,  
Poincar\'e gauge theory, where 
a symmetric stress-energy tensor is only possible for vanishing spin. 
 
For the simplified equations (56)-(57), the reason for the necessity 
of the mass term (i.e., the $c$-terms) is especially clear. For 
$c=0$, we are dealing with an Einstein-Maxwell system. Thus, the torsion 
will only be determined up to a gauge transformation $A_m \rightarrow A_m 
+ f_{,m}$. However, measurable  results, like the evolution of spin and 
momentum of a test particle in the given field   configuration \cite{5}, 
will depend on the full torsion tensor. Otherwise stated, 
the field equations for $c=0$ do not completely determine the physical 
 fields.  

Unfortunately, the mass term, necessary on theoretical grounds, is 
rather disturbing from a computational point of view, since the  
known solutions of the Einstein-Maxwell system do not easily generalize 
to the Proca field.

\section{Discussion and solutions of the field equations}

In the last section, we have reduced the field equations of our  
specific Poincar\'e gauge theory  to a purely Riemannian  
problem: The description of a massive vector field in curved spacetime. 
The discussion of the system (56)-(57) is therefore subject to 
the framework of general relativity, and can be found in the corresponding 
literature. Nevertheless, for the sake of completeness, we will 
briefly sketch the main features of this system 
and refer to 
the original articles for details. 

It is well known that 
in the cases of practical interest, no exact solution of the 
Einstein-Proca field equations has been found 
to date. Especially, no generalization of the spherically symmetric 
Reissner-Nordstr\o m solution, nor of the axially symmetric Kerr-Newman 
solution are known for the case of a massive vector field. Moreover, 
it has been established \cite{9} that there is no black hole solution 
other then for $A_i = 0$. This is a special case of the no-hair 
theorem for black holes. Here, we are interested in neutron stars. 
In view of the no-hair theorem, the vacuum solutions we eventually find for 
the exterior of the star will appear with a  naked singularity at the 
origin. This, however, does not mean that they are unphysical. As vacuum  
solutions, they are restricted to the exterior of the star and will have 
to be completed by a suitable interior solution, thereby removing the 
singularity. On the other hand, the absence of black hole solutions 
 indicates that there is a rather fundamental difference between 
the Maxwell and the Proca case. Especially, it seems that there are no 
solutions   
that tend to the Einstein-Maxwell black holes (Kerr-Newman or 
Reissner-Nordstr\o m) for $c \rightarrow 0$ ($c$ is the mass parameter of 
the vector field in (56)-(57)), as might have been expected.   

A detailed  analysis of the static,  spherically symmetric 
case has been performed and numerical  solutions 
have been derived in Refs. \cite{10}-\cite{12}. The authors of the three 
articles essentially agree in their conclusions, and there is no need 
to repeat the analysis here. The main feature of the solutions is that, 
for large distances from the 
source, the metric tends to the Schwarzschild solution. This is 
in agreement with  
the expected exponential decrease of the massive vector field known from 
the special relativistic limit. Indeed, for $g_{ik} = \eta_{ik}$, 
 starting with the general spherically symmetric, static 
field  $A_i = (\phi(r), \psi(r) \frac{\vec x}{r})$ in the comoving 
system of the 
source,  $u^i = \delta^i_0$, one easily 
shows from (56) that $\psi(r) = 0$, and that $\phi(r)$ is 
solution to the  equation 
\begin{equation}
- 4 b \Delta \phi + 2c \phi = \rho, 
\end{equation}
with solution 
\begin{displaymath}
\phi \sim \frac{e^{- \sqrt{c/2b}\ r}}{r}. 
\end{displaymath}
In order to fix the proportionality factor, we have to take a look at  
equations 
(50) and (51). In contrast to the usual form of the Maxwell (or Proca) 
equations, 
the \textit{charge} $\sigma_{ab}$ has essentially been 
removed from $A_i$ in the ansatz for  $K^{ab}_{\ \ m}$, i.e., 
in the flat case, the density $\rho(r)$ is supposed to be normalized 
to $\int \rho(r) \de^3 x = 1$. It is then easy to see (consider the 
special case 
$c=0$ and $\rho(r) = \delta(r)$) that the complete solution reads 
\begin{equation}
\phi(r)=  \frac{1}{16 \pi b }\  \frac{e^{- \sqrt{c/2b}\ r}}{r}. 
\end{equation}

Since the systematic numerical analysis can be found in Refs. 
\cite{10}-\cite{12}, 
we will apply here a different approach to the system (56)-(57). Namely, 
we will proceed  in the spirit of the WKB approach of quantum mechanics,   
which consists in expanding the system in terms of Planck's constant. 
In our case, this means expanding in terms of the spin tensor $\sigma_{ab}$ 
or simply in terms of the parameter $s$ introduced in (54). 
To 
zeroth order (i.e., for $s=0$), we 
simply get the Schwarzschild solution with $A_i = 0$. 
We are interested in the first order corrections to this solution. 
Thus, in (56)-(57), we simply neglect the terms in $s^2$. 
Again, form (57), we find the Schwarzschild solution. (Just as in 
Einstein-Maxwell theory, where the corrections to the metric are 
of second order in the electric charge, here, the corrections are 
of second order in the spin.) At this stage, we are dealing with 
a Proca field on a Schwarzschild background. Finally, in order 
to solve equation (56), we expand the metric in terms of the 
Schwarzschild mass $m$ and neglect the term of order $\phi(r) \frac{m}{r}$
and higher, leaving us with the solution (59).  

Summarizing, at large distances, our solution looks like the 
Schwarzschild solution with $A_i=0$. Approaching the source, 
the first order corrections are given by the solution (59) for 
$A_i = (\phi, 0,0,0)$. The next order of approximation consists 
in interaction terms of the gravitational potential $m/r$ and the 
vector field $A_i$, leading to corrections in (59). Only at the 
next order ($\sim A^2$) will the corrections to the Schwarzschild 
metric become apparent. 

Note that the Schwarzschild metric  is not only a very good approximation 
 because of the exponential decrease of $\phi$, but also because of the 
absolute smallness of spin effects in general. In practice, rotational 
effects (leading to Kerr type corrections) will by far dominate the 
(intrinsic) spin effects. Not only will the rotational spin of a neutron star 
be much larger than its intrinsic spin, but moreover, the rotational 
effects lead to first order corrections in the metric, in contrast to the 
second order spin effects. 
 Experimentally, this means that the spin effects cannot 
be observed by analyzing the geodesics, but rather by measuring directly 
the torsion field (through its interaction with  spinning test bodies, see 
Ref. \cite{5}). 

Let us emphasize  that, although the Schwarzschild metric 
is subject to higher order corrections, it will retain its spherical 
symmetry to each order. The same holds for $A_i$, but not for 
$K^{ab}_{\ \  i}$, which is given by (51).  

Unfortunately, the analysis of Refs. \cite{10}-\cite{12} has not been 
extended to the axially symmetric case. Since this is the case 
of practical interest (neutron stars are supposed to be subject to 
a rotation of very high frequency), we will extend the simple 
approach of the above considerations also to this case. As before, 
we  thus  neglect terms of second order in the spin. This, clearly, 
leads to the Kerr metric. Then, we also neglect interactions of the 
gravitational potential with the vector field $A_i$. 

First, let us recall the Kerr-Newman solution for the massless vector 
field, i.e., the axially symmetric solution of (56)-(57) for $c=0$. 
Using spherical coordinates $x^i = (t,r,\theta,\phi)$, 
the Kerr-Newman metric reads 
\begin{eqnarray}
 \  \ \ \ \ \ \ \ \ \  
\de s^2 = \frac{\rho^2 - 2mr + q^2}{\rho^2} \de t^2 
- \frac{\rho^2}{\Delta} \de r^2 -\rho^2 \de \theta^2 \nonumber \\ 
- \frac{(2mr - q^2) 2 a \sin^2 \theta}{\rho^2} \de t \de \theta
-[a^2 + r^2 + a^2 \sin^2 \theta \  \frac{2mr - q^2}{\rho^2}] \de \phi^2, 
\end{eqnarray}
with $\rho^2 = r^2 + a^2 \cos^2 \theta,\ \Delta = r^2 - 2mr + q^2 + a^2$. 
Here, $m,q,a$ are constants of integration. As is easily seen taking 
suitable limits, $m$ is the Schwarzschild mass, $a$ is related to the 
angular momentum of the source (see any textbook on general relativity) 
and the charge parameter $q$ is proportional to the 
intrinsic spin $s$ (see below). 
The corresponding field $A_i$ has the form 
\begin{displaymath}
A_i = \frac{q} {s \sqrt{2 b}}\ ( \frac{r}{\rho^2}, 0, 0, - \frac{ a r 
\sin^2 \theta}{a^2 \cos^2 \theta + r^2 }).
\end{displaymath}
The constant $q$ can be evaluated by taking the Coulomb limit, i.e., 
take $a = 0$ and compare with Eq. (59) for $c=0$. The result is 
\begin{equation}
q =  \frac{s\sqrt{2}}{16 \pi \sqrt{ b}}, 
\end{equation}
and the solution takes the form 
\begin{equation}
A_i =  \frac{1} {16 \pi  b}\ ( \frac{r}{\rho^2}, 0, 0, - \frac{ a r 
\sin^2 \theta}{a^2 \cos^2 \theta+ r^2}).
\end{equation}
This is the exact solution for the massless vector field. A difference to 
the spherically symmetric case arises if we consider the flat limit 
of this solution. In the Reissner-Nordstr\o m case, the solution $A_i$ is 
simply the Coulomb potential, which is a solution of $F^{ik}_{\ \ ;i}=0$
in the flat case too. At first sight, it seems as if (62) is not a 
solution on the flat background $\de s^2 = \de t^2 - \de r^2 - r^2 (\de
\theta^2 + \sin^2 \theta \de \phi^2)$. This, however, is the result 
of a false interpretation of the coordinate system. Indeed, the flat limit 
of (60) is found for $q \rightarrow 0, m \rightarrow 0$, i.e., 
\begin{displaymath}
\de s^2\!= \! \de t^2\! - \frac{r^2 + a^2 \cos^2 \theta}{r^2 + a^2} 
\de r^2 \!-\! (r^2+a^2 \cos^2 \theta) \de \theta^2 
\!-\! (a^2 + r^2 )\sin^2 \theta \de \phi^2, 
\end{displaymath}
which is easily shown to be flat ($\hat R^i_{\ klm} = 0$) and 
related to the Minkowski metric $\eta_{ik} = diag (1, -1,-1,-1)$ by 
the coordinate transformation 
\begin{eqnarray*}
x &=&  \sqrt{r^2 + a^2} \sin \theta \cos \phi \\
y &=& \sqrt{r^2 + a^2} \cos \theta \cos \phi \\
z &=& r \cos \theta. 
\end{eqnarray*}
We now find that (62) is indeed a solution in the flat background 
expressed in the non-spherical coordinates $r, \theta, \phi$. Since 
it is rather difficult to generalize this to the massive case, 
we will neglect contributions that are of second order in the 
rotational momentum $a$. To that order, the coordinates used in (62) 
coincide with conventional  spherical coordinates, and indeed 
the field 
 \begin{equation}
A_i =  \frac{1} {16 \pi  b}\ ( \frac{1}{r}, 0, 0, - \frac{ a  
\sin^2 \theta}{ r}), 
\end{equation}
which differs from  (62)  by terms of order $a^2$ and higher, 
is a solution to the Maxwell equations on a flat background (in 
spherical coordinates). Physically, we see that (63) contains an 
electric monopole and a magnetic dipole contribution, while (62) 
contains, as is well known, the complete series of odd electric 
multipoles (monopole, quadrupole\dots ) 
and even magnetic multipoles (dipole, octupole\dots ). 
Thus, replacing (62) by (63), 
we essentially neglect quadrupole and higher moments. 
We then make the following ansatz for the massive vector field:
\begin{equation}
A_i =  \frac{1} {16 \pi  b}\ ( \frac{1}{r} f(r), 0, 0, - \frac{ a  
\sin^2 \theta}{ r}g(r)). 
\end{equation}
Putting this into (56) yields, on a flat background, 
 the unique asymptotically flat solution 
\begin{eqnarray}
f(r) &=& e^{-\sqrt{c/2b}\ r} \nonumber \\ 
g(r) &=& (1+ \sqrt{c/2b}\ r) e^{-\sqrt{c/2b}\ r}, 
\end{eqnarray} 
where the constants of integration have been fixed by requiring the 
limit (63) for $c \rightarrow 0$. 

Summarizing, the Kerr metric, as before the Schwarzschild metric, 
is a very good approximation for three reasons: (1) The spin is 
small compared to the angular momentum of the source. (2) Eventual  
corrections to the metric are of second order in $A_i$ and thus in 
 the spin (in contrast to the angular momentum contributions which are 
first order). (3) The field $A_i$, and thus the metric 
corrections, decreases exponentially. 

The first order approximation of $A_i$ is given by (64)-(65), neglecting
 interactions of the metric with $A_i$ (order $\sim A_i\ m/r$) and 
second order terms in the spin and in the rotational momentum. 

Possibly, one can find exact solutions of (56) in Schwarzschild 
or Kerr backgrounds. This would yield the complete analogue of the 
WKB approximation, 
neglecting only terms of second and higher order in the spin, but 
including the spin-gravity couplings $\sim A_i\ m/r$. 
 However, as we have argued before, it 
is not the scope of this article to analyze in detail the equations 
 in the form (56)-(57), because this set of equations is well known 
from classical general relativity. 

Finally, one could also consider the possibility of a very small 
$c$, too small for the exponential decrease to be sensible on 
astrophysical scales. (In analogy to theories with a small photon 
mass.) In this case, one can consider the term in $c$ as a gauge fixing term 
but irrelevant else. This means that we can simply take over the solutions 
from Einstein-Maxwell theory, i.e., the Reissner-Nordstr\o m metric together 
with the Coulomb potential, or the Kerr-Newman metric with the field 
(62), but in the gauge that is consistent with $c \neq 0$, which is 
essentially $A^i_{;i} =0$, as can be easily derived taking the 
divergence of (56). Note that (62) and also the Coulomb solution in its 
usual form are already in this gauge. Thus, the term in $c$ forces 
us to choose the gauge $A^i_{;i} =0$, but the $c$-contributions 
will then be neglected in the field equations. 

Does every solution of (56)-(57) tend to a solution of the Einstein-Maxwell 
equations in the limit $c =0$? Our approximate solutions seem to indicate 
that this is indeed the case. One should, however, be cautious, 
because there are nevertheless severe differences between the massive and 
the massless vector fields. One example is the previously mentioned 
absence of black hole solutions in the massive case. Thus, there might be 
 surprises in the strong field region which have not been revealed by 
our  approximations. 

Let us remind that the whole section was based on the specific 
ansatz (51), which was adapted to the macroscopic spin distribution 
(50). This does not mean that, even if we retain (50), there are no 
other solutions. Especially, the existence of black holes cannot 
be excluded a priori.  
On the other hand, for a different spin density, 
for instance the totally antisymmetric tensor of the Dirac 
particle, a different ansatz will be necessary, leading to 
entirely different solutions.

\section{Spin precession in torsion field} 

Finally, we address the question of the detectability of 
the torsion fields by analyzing the behavior of test bodies 
in a given field configuration. We confine ourselves to the 
concrete solutions of the previous section. 

It is well known that the equations for the spin precession and 
for the momentum evolution 
in a Riemann-Cartan geometry depend on the nature of the test 
body in question. Especially, elementary particles with different 
spin couple in a different manner to the torsion. A complete 
review, as well as a simple method for the derivation of the 
equations of motion, has been presented in Ref. \cite{5}, where the 
references to the original articles can be found.    

For simplicity, 
we consider again the case where the test particle is a macroscopic, 
spin polarized body. Thus, the system (central source + test body) we are 
interested in is essentially a neutron star binary, under the 
restriction, for computational simplicity, that the central star has a 
much larger mass and can be considered to be at rest. 

The deviation from geodesic motion is given by a direct coupling  
of the test particles' spin to the curvature tensor. Even for 
the classical spin (i.e., the 
angular momentum in the body's rest frame), which presents a similar 
coupling to the curvature (Papapetrou equations in general relativity),
these corrections have not yet been measured in experiment,  
and the contributions from the intrinsic spin and the non-Riemannian 
fields will be even smaller. Therefore, the only experimentally relevant 
equation is the spin precession equation, which reads \cite{5}
\begin{equation}
\De \sigma^{(2)}_{ik} + \hat \De S^{(2)}_{ik} = P_i u_k - P_k u_i,  
\end{equation}
where $\De $ denotes the covariant derivative with respect to proper 
time using the full connection $\Gamma^i_{lm}$ and $\hat \De $ 
the corresponding derivative with the Christoffel connection, i.e., 
$\De \sigma^{(2)}_{ik} = \de \sigma^{(2)}_{ik}/\de \tau - \Gamma^l_{im} 
\sigma^{(2)}_{lk}u^m 
- \Gamma^l_{km}\sigma^{(2)}_{il}u^m$ and  $\hat \De S^{(2)}_{ik} 
= \de S^{(2)}_{ik}/\de \tau - \hat \Gamma^l_{im} S^{(2)}_{lk}u^m 
- \hat \Gamma^l_{km} S^{(2)}_{il}u^m$. We denote by $\sigma^{(2)}_{ik}$ 
the (intrinsic) spin tensor of the test body and by $S^{(2)}_{ik}$ 
its angular momentum in the rest frame (due to rotation). The index 
$(2)$ is attached to avoid confusion with the corresponding quantities of the 
source. $P_i$ is the momentum vector of the test body, which is not 
necessarily parallel to the velocity $u^i$. However, 
the right hand side of (66) is of higher order \cite{5} and will be 
neglected in the following. Finally, we have to impose some conditions 
on the structure of the test body. It seems reasonable, for neutron stars, 
to make the ansatz (strong spin-rotation coupling) 
\begin{equation}
S^{(2)}_{ik} = g \sigma^{(2)}_{ik}, 
\end{equation}
which leaves us with 
\begin{equation}
\hat \De \sigma^{(2)}_{ik} + \frac{1}{1+g}K^l_{\ im}\sigma^{(2)}_{lk} u^m 
+ \frac{1}{1+ g} K^l_{\ km}\sigma^{(2)}_{il} u^m = 0,  
\end{equation}
where we have separated into Riemannian and  non-Riemannian contributions.   
 Next, introduce the spin vector $\sigma^i_{(2)} = \frac{1}{2} \eta^{iklm}
u_k \sigma^{(2)}_{lm}$ ($\eta^{iklm} = |g|^{-1/2} \epsilon^{iklm}$) 
and use the spin supplementary condition 
 $\sigma^{(2)}_{ik}u^k = 0$ to find 
\begin{equation}
\hat \De \sigma^i_{(2)} + \frac{1}{1+g} K^i_{\ lm}\sigma^l_{(2)}u^m = 0,  
\end{equation}
where we have used the approximate validity of the geodesic equation 
$\hat \De u^i =0$ and higher order terms have been omitted \cite{5}. 
This equation, together with the  solutions of the previous section, 
can be used to evaluate the spin precession and especially to 
measure the influence of the torsion contributions. 

Here, our scope is to illustrate the effects of the dynamical torsion 
fields. Therefore, we confine ourselves to 
the simple case of the non-rotating test body ($S^{(2)}_{ik} 
= 0$, i.e., $g=0$) and use the spherically symmetric solution 
(no rotation of the source). This is not a 
very realistic case (as we have argued before, the rotational 
effects, when dealing with neutron stars,  
will in general be by far more important than the intrinsic spin
effects), but it is a useful example to illustrate directly the 
post-general relativistic effects due to  spin and torsion. 

With $K^{ab}_{\ \ l}$ from (51), with $A_m = (\phi, 0,0,0)$ 
and introducing the spin vector of the source,  
$\sigma^i_{(1)} = \frac{1}{2} \eta^{iklm}u_k \sigma^{(1)}_{lm}$,  
using only the Newtonian order of the Schwarzschild background 
(compare with Ref. \cite{5}), we find to lowest order for the relevant 
spin precession equation
\begin{equation}
\frac{\de \vec \sigma^{(2)}}{\de t} = \frac{3}{2}\ \frac{m}{r^3} 
 [\vec L \times \vec \sigma^{(2)}] + \phi(r) [\vec \sigma^{(1)} \times 
\vec \sigma^{(2)}], 
\end{equation}
where $\phi(r)$ is given by Eq. (59). The first term, with the orbital 
momentum $\vec L$, is identical to the precession of the rotational 
spin in general relativity (as derived from the lowest approximation 
of the Papapetrou equations). 
This is a general result of Poincar\'e gauge theory: In purely 
Riemannian spacetimes, no distinction can be made between intrinsic 
and classical (rotational) spin. The torsion, however, couples only 
to the intrinsic spin. Therefore, the other way around, only a particle 
with intrinsic spin can distinguish between Riemannian and non-Riemannian 
geometry. 

The above result is easily generalized to the more realistic case 
of rotating neutron stars. One can also proceed to a more complete 
treatment considering the neutron star binary as a two body problem. 
On the other hand, Eq. (70) is already enough to discuss the 
possibility of an experimental detection. It is clear that the 
spin effects are very small, even for  completely spin polarized stars. 
Moreover, the torsion decreases exponentially, and practical results 
will strongly depend on the coupling constants $b$ and $c$, especially on 
$c$.  A possible, indirect, detection could  be based on gravitational 
waves emitted by a merging binary system, since this will allow 
for conclusions on the evolution of the system in the strong field regime.

\section{Conclusions}

Considering, in the framework of Poincar\'e gauge theory,  
the Christoffel part of the Lorentz connection as a function 
of the tetrad field in a formalism similar to the second order formalism 
used in general relativity and in metrical theories in general, we were 
able to present a theory that allows for dynamically propagating torsion 
fields, preserving nevertheless a full general relativity limit in the 
absence of spinning matter. This is not possible in the framework of the 
conventional, first order formalism, without introducing higher order 
derivatives of the independent field variables and apparently 
running into trouble 
with the Cauchy initial value problem. The  equivalence of  
the second order formalism with the conventional approach has been established 
and therefore, ultimately, we have 
shown that in the conventional formalism, certain Lagrangians lead to 
consistent theories despite the appearance of higher derivatives.  
A concrete model has been presented, and for  a specific class of solutions, 
the field equations have been reduced to an Einstein-Proca system 
using a slight modification of the Yasskin method known from conventional 
Einstein-Yang-Mills theories. Approximate solutions have been 
discussed for concrete cases, corresponding to the exterior of spin 
polarized neutron stars. In view of an eventual experimental detection 
of the non-Riemannian effects, the spin precession equation of a test 
body in such a field configuration has been briefly analyzed. 

\section*{Acknowlegments}

This work has been supported by EPEAEK II in the framework
 of ``PYTHAGORAS
II - SUPPORT OF RESEARCH GROUPS IN UNIVERSITIES'' (funding: 75\% ESF - 25\%
National Funds).

\end{document}